\documentclass[twocolumn,preprintnumbers,amsmath,amssymb]{revtex4}
\usepackage{graphicx}
\usepackage{dcolumn}
\usepackage{bm}

\begin{document}
\title{Non-additive dynamical Casimir atomic phases }
\author{Fran\c{c}ois Impens$^{1}$, Claudio Ccapa Ttira$^{2}$, and Paulo A. Maia Neto$^{2}$}
\address{$^{1}$ Observatoire de la C\^{o}te d'Azur (ARTEMIS), Universit\'e de Nice-Sophia Antipolis, CNRS, 06304 Nice, France}
\address{$^{2}$ Instituto de F\'{i}sica, Universidade Federal do Rio de Janeiro,  Rio de Janeiro, RJ 21941-972, Brazil}
\begin{abstract}
We discuss a fundamental property of open quantum systems: the quantum phases associated with their dynamical evolution are non-additive. We develop our argument by considering a multiple-path atom interferometer in the vicinity of a perfectly conducting plate. The coupling with the environment induces dynamical corrections to the atomic phases. In the specific example of a Casimir interaction, these corrections reflect the interplay between field retardation effects and the external atomic motion. Non-local open-system Casimir phase corrections are shown to be non-additive, which follows directly from the unseparability of the influence functional describing the coupling of the atomic waves to their environment. This is an unprecedented feature in atom optics, which may be used in order to isolate non-local dynamical Casimir phases from the standard quasi-static Casimir contributions.
\end{abstract}
\maketitle

Open quantum systems~\cite{CalzettaHu,Petruccione} have motivated a worldwide theoretical and experimental research effort. Basic quantum phenomenon such as decoherence~\cite{DecoherenceReview} have been reported in a variety of mesoscopic systems. Atom interferometers~\cite{Cronin09} in the vicinity of a conducting surface constitute a particulary relevant and rich class of open quantum systems, in which both long-lived (atomic dipole) and short-lived (electric field) degrees of freedom are simultaneously at work. 

Here, we propose to use this example in order to demonstrate an unprecedented, key property of open quantum systems: the non-additivity of the quantum phases arising from their dynamical evolution. The coupling to the environment is described by an influence functional~\cite{FeynmanVernon} depending simultaneously on a pair of quantum paths. This stands in sharp contrast to the quantum phases resulting from a unitary evolution, which depend only on single paths taken separately. While the phase differences associated to single-path contributions are additive by construction, the additivity has no reason to be valid for the influence functional phases associated to pairs of paths. In general, the double-path influence functional phases cannot be separated into sums of single-path contributions~\cite{CalzettaHu}. The non-additivity of the environment-induced quantum phases is a direct consequence of this unseparability, which is intimately connected to the non-locality of these phases. 

 Atom interferometers have been used to probe atom-surface interactions in the van der Waals (vdW) regime~\cite{Perreault05,Lepoutre09}, turning atom optics into a promising field for the experimental investigation of dispersive forces~\cite{Dimopoulos03,Carusotto05,Derevianko09,Wolf07,Pelisson12,Zhou13}. The effect of surface interactions onto atomic waves propagating near a conducting plate is commonly described by means of the vdW (or Casimir-Polder at longer distances) potential taken at the instantaneous atomic position. In this description, the external atomic waves are treated as a closed quantum system driven by conservative forces. 
 
 Nevertheless, we have shown recently~\cite{DoublePath} that such an approach is incomplete. This is so, because the external atomic degrees of freedom (d.o.f.s) are coupled to the internal dipole and electromagnetic field fluctuations. 
Thus, the atomic waves propagating in the vicinity of a conducting surface behave essentially as an open quantum system coupled to an environment (the dipole and electromagnetic field d.o.f.s)~\cite{Ryan10a,Ryan10b,Ryan11}. In addition to the expected decoherence \cite{Stern90,Anglin03,Mazzitelli03,Sonnentag07,Scheel12}, the non-unitary evolution of the atomic waves gives rise to specific atomic phase shifts. A similar real  phase shift has been discussed in the context of geometrical phases in spin-boson systems coupled to an environment~\cite{Whitney05,Lombardo06}.

In this article, we use the formalism 
 of Ref.~\cite{DoublePath} in order to obtain explicit results for the dynamical corrections to the Casimir phase for general atomic trajectories close to
  a conducting plate. We show that the local dynamical corrections can be captured by coarse-graining the Casimir potential over a finite time-scale. 
 More importantly, we demonstrate that 
 the non-local double-path dynamical phase contributions
 are non-additive, illustrating a key property of the non-unitary evolution of open quantum systems.
 The phase additivity can only be discussed for an atom interferometer geometry with at least three distinct paths, as depicted in Fig.~1. Thus, our first step will be to generalize the approach of Ref.~\cite{DoublePath} to multiple-path atom interferometers~\cite{Zhou13,Weitz96,Hinderthur99,Impens09b,Robert10,Impens11} evolving nearby a perfect conductor.

In standard atom interferometry, one associates well-defined phases to individual paths, and as a consequence we are allowed, for instance, to add phase coherences between 
    arms 1 and 3 and 3 and 2 to find the phase coherence between arms 1 and 2. However, this is no longer the case when taking the non-local corrections into account, because the non-local double-path (DP) phase coherences are associated to pairs of paths rather than to individual ones. In the specific case of van der Waals atom interferometry, the non-additive phases appear in the form of a non-local relativistic correction to the standard atomic van der Waals phases. In contrast to other local relativistic corrections, also discussed in this letter, non-local relativistic corrections may be isolated from the much larger quasi-static contributions thanks to their non-additivity.

The three  paths propagating in the half space $z >0$ interact between $t=0$ and $t=T$ with a nearby  perfectly conducting plate located in the plane $z=0$ as shown in Fig.~1. In addition to the dispersive interaction with the plate, 
atoms are driven by an external potential $V_{\rm ext}(\mathbf{r})$, linear or quadratic in position. 
We assume the atomic motion non-relativistic. 
 The atomic state is initially a sum of three Gaussian wave-packets with negligible overlap, $| \psi(0) \rangle = 
 \frac{1}{\sqrt{3}}\sum_{k=1}^3 |\psi_k(0) \rangle$, with $\langle \mathbf{r} | \psi_k(0) \rangle = wp (\mathbf{r},\mathbf{r}_{0 \:k},\mathbf{p}_{0\: k},\mathbf{w}_{0 \: k}) $. These packets have a central position $\mathbf{r}_{0 \:k}$, momentum $\mathbf{p}_{0\: k}$ and a width $\mathbf{w}_{0 \: k}:$
  $wp(\mathbf{r},\mathbf{r}_{0},\mathbf{p}_{0},\mathbf{w}_0)  =  \prod_{\eta=x,y,z} (1/\sqrt{\pi \: w_{0 \eta}}) e^{- (\eta-\eta_{0})^2 /2 w_{0 \eta}^{2}  + i  p_{ 0 \eta} (\eta-\eta_{0}) / \hbar}$. We consider a sufficiently dilute sample, so that atomic interactions effects can be neglected.
\begin{figure}[htbp]
\begin{center}
\includegraphics[width=6cm]{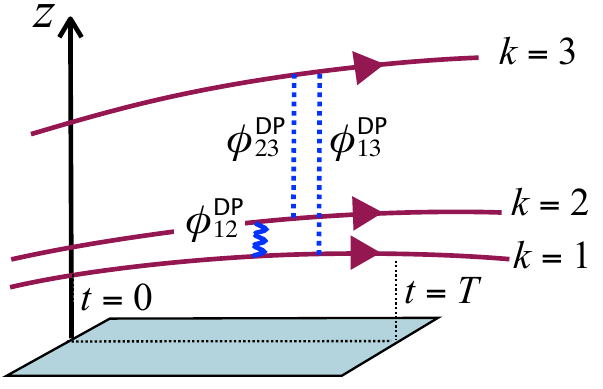}
\end{center}  \caption{(color
  online).
  Three-arm atom interferometer near a conducting plate at $z=0$. When the arm $k=3$ is further away from the plate, the non-local double-path phases 
  $\phi_{23}^{\rm DP}$ and  $\phi_{13}^{\rm DP}$ are much smaller than $\phi_{12}^{\rm DP}$ thus enhancing the 
  non-additivity effect.  }
\label{fig:three arms atom interferometer}
\end{figure} 

 We first present the standard analysis of this atom interferometer by means of an instantaneous vdW potential $V_{{\rm vdW}}(\mathbf{r})$.   Provided that the vdW potential on the atoms is weak enough so as to make dispersion effects negligible, an excellent approximation in the experimental conditions of Ref.~\cite{Perreault05,Lepoutre09}, one can apply the $ABCD$ propagation method~\cite{Borde01,Borde02,Impens09a} for atomic waves in quadratic potentials: at any time $t>0$, each
  atomic wave-packet is given by 
  \begin{equation}\label{ABCD}
  |  \psi_k(t) \rangle = | \chi_k(t) \rangle e^{i [\varphi_k^{(0)}(t)+  \varphi_{k}^{({\rm vdW})}(t)]} 
 \end{equation} 
   with a time-dependent Gaussian $\langle \mathbf{r}|  \chi_k(t) \rangle  =  wp(\mathbf{r},\mathbf{r}_k(t),\mathbf{p}_k(t),\mathbf{w}_k(t))$. The precise value of the width vector $\mathbf{w}_k(t)$ is not important for the coming discussion. The average atomic position $\mathbf{r}_k(t)$ and momentum $\mathbf{p}_k(t)$ follow the classical equations of motion with the initial conditions $\mathbf{r}_k(0)=\mathbf{r}_{0 \:k}$ and $\mathbf{p}_k(0)=\mathbf{p}_{0 \:k}$ associated with the central trajectory 
  corresponding to path $k$ 
  ($k=1,2,3$).  More important for our discussion are the phase contributions in Eq.~(\ref{ABCD}). 
  The phase $\varphi_k^{(0)}(t)$ 
  collects the free propagation and external potential effects, whereas 
 $ \varphi_{k}^{({\rm vdW})}(t)$ accounts for the dispersive atom-surface interaction.
  From now on, we focus on the phase accumulated between the instants $t=0$ and $t=T$, omitting explicit reference to time $T$ to alleviate notations. 
  The phase $\varphi_{k}^{(0)}$ is given by the following integral along the trajectory $k$:
   $\varphi_{k}^{(0)}= \frac {1} {\hbar} \int_{0}^{T} \! dt \left(  \frac {\mathbf{p}_k^2(t)} {2 m} - E(t) - V_{\rm ext}(\mathbf{r}_k(t)) \right)$,  where $E(t)$ is the internal atomic energy at time $t$. In this standard approach, the atom-surface interaction simply yields an additional phase shift given by the integration of the  vdW potential $ V_{{\rm vdW}}(z)$ taken at the instantaneous atomic positions  
along the path  $k:$
 \begin{equation}
 \label{eq:instantaneous vdW potential phase}
  \varphi_{k}^{({\rm vdW})}= -\frac 1 \hbar \int_0^T dt\, V_{{\rm vdW}} (z_k(t))
  \end{equation}
  
The density matrix corresponding to the atomic state at time $T$ computed within the standard \textit{ABCD} approach is then given by
\begin{eqnarray}
 \label{eq:evolved density matrix nonunitary}
{\rho}(T)  =  {\rho}_{\rm diag}(T)  + \frac 1 3
\left(\sum_{j<k}^3  | \chi_j(T) \rangle \langle \chi_k(T)  | e^{i \phi_{jk}^{\rm st}}
 + \mbox{H.c.}\right)
 \end{eqnarray}
with  ${\rho}_{\rm diag}(T)\equiv  \frac 1 3\sum_k | \chi_k(T) \rangle \langle \chi_k(T) | $ and $ \mbox{H.c.}$ representing the Hermitian conjugate.
 We focus here on the standard phase coherences $\phi_{jk}^{\rm st}$, 
 \begin{eqnarray}
  \phi_{jk}^{\rm st}&=& \phi_{jk}^{(0)} +\varphi_{j}^{({\rm vdW})}-\varphi_k^{({\rm vdW})},\\
 \phi_{jk}^{(0)}&\equiv&\varphi_{j}^{(0)}-\varphi_k^{(0)}.
 \end{eqnarray}
They (obviously) satisfy  additivity:
 \begin{equation}
\label{eq:additivity property}
\phi_{jk}^{\rm st} =\phi_{j\ell}^{\rm st}+\phi_{\ell k}^{\rm st}
 \end{equation}
 for any $j,k,\ell=1,2,3,$ since they originate from phases associated to individual paths in Eq.~(\ref{ABCD}). 
 
 We now analyse the multiple-path atom interferometer as an open quantum system, building on our recent work~\cite{DoublePath}, and 
 show that the additivity condition (\ref{eq:additivity property}) no longer holds.  
 We start from the full quantum system,
 whose dynamics is described by the Hamiltonian $\hat{H}=\hat{H}_E+\hat{H}_D+\hat{H}_F+\hat{H}_{AF},$
 including the external  ($\hat{H}_E$), internal ($\hat{H}_D$) and electromagnetic field ($\hat{H}_F$) d.o.f.s.
 The interaction Hamiltonian, which reads in the electric dipole approximation $\hat{H}_{AF} = - \hat{\mathbf{d}} \cdot \hat{\mathbf{E}}(\hat{\mathbf{r}}_a)$, couples the  atomic center-of-mass $\hat{\mathbf{r}}_a$  to the internal dipole $\hat{\mathbf{d}} $ and  the electric field $\hat{\mathbf{E}}$.

 The external atomic waves are described by the reduced atomic density matrix obtained after coarse-graining over the field and internal atomic d.o.f.s. These play the role of an environment, whose effect on the atomic waves is captured by an influence phase $S_{\rm IF}[\mathbf{r}_j,\mathbf{r}_k]$~\cite{DoublePath}:
\begin{eqnarray}
 \label{eq:evolved density matrix nonunitary}
{\rho}(T) &  = & {\rho}_{\rm diag}(T)  + \\
&&  \frac 1 3
\left(\sum_{j<k}^3  | \chi_j(T) \rangle \langle \chi_k(T)  | e^{i (\phi_{jk}^{(0)}+ \frac {1} {\hbar} S_{\rm IF}[\mathbf{r}_j,\mathbf{r}_k])}
 +\; \mbox{H.c.}\right)\nonumber
 \end{eqnarray}

 The complex influence phase $\frac {1} {\hbar }S_{\rm IF}[\mathbf{r}_j,\mathbf{r}_k]$, evaluated along the central atomic trajectories $j$ and $k$ 
  (a valid approximation for narrow wave-packets), describes completely the atom-surface interaction effects. Its imaginary part corresponds to the plate-induced decoherence, and its real part  gives the atomic phase shift arising from surface interactions. This phase contains local contributions involving a single path 
   (SP) at a time, and a non-local double-path (DP) contribution involving simultaneously two paths: 
 \begin{equation}\label{IF}
\frac {1} {\hbar} \mbox{Re} \left[ S_{\rm IF}[\mathbf{r}_j,\mathbf{r}_k] \right]
= \varphi^{\rm SP}_j-\varphi^{\rm SP}_k+\phi^{\rm DP }_{jk}.
 \end{equation}

In this letter, we  provide explicit analytical results for the single  and double-path phase contributions 
 in the short-distance van der Waals limit $\omega_0 z_k/c\ll 1,$ which yields larger 
 phase shifts and
  matches the conditions of the experiments performed so far~\cite{Perreault05,Lepoutre09}. 
In this regime, the dominant contribution comes from the symmetric dipole correlation function $\langle \{{\hat d}(t),{\hat d}(t')\}\rangle$ (${\hat d}$ is any 
Cartesian component of the vector operator $\hat{\mathbf{d}}$), which contains the information about the 
quantum dipole fluctuations,
whereas the symmetric electric field correlation function yields a negligible contribution. 

In the short-distance limit,
the relevant field correlation function is 
 the  retarded Green's function representing the electric field linear response susceptibility to 
the fluctuating dipole source:
\begin{equation}
{\cal G}^R_{\hat E}({\bf r},t;{\bf r}',t')\equiv  \frac{i}{\hbar} \theta(t-t')\sum_{\eta=x,y,z}\langle [{\hat E}_{\eta}({\bf r},t),{\hat E}_{\eta}({\bf r}',t')]\rangle,
\end{equation}
with $\theta(t)$ representing the Heaviside step function. 
${\cal G}^R_{\hat E}({\bf r},t;{\bf r}',t')$ is the sum of two contributions: 
 the free space 
Green's function, which represents the direct propagation from  ${\bf r}'=(x',y',z')$ at time $t'$ to  $\bf r$ at time 
$t$ 
 and does not contribute 
to the surface interaction; and the scattered Green's function ${\cal G}^{R,S}_{\hat E},$ which accounts for 
the propagation containing one reflection at the surface~\cite{Wylie84,Wylie85}. The latter is written in terms of the source point image ${\bf r}'_I=(x',y',-z'),$ 
represented in Fig.~2, and vanishes outside the light cone defined by the condition $\tau\equiv t-t'=|{\bf r}-{\bf  r'_I}|/c.$

We first address the local single-path phases $\varphi^{\rm SP}_k$ in (\ref{IF}). 
From the general expression for the SP phase given in \cite{DoublePath}, we find in the short-distance limit
\begin{eqnarray}
 \varphi^{\rm SP}_{k} \!  =  \! \frac {1} {4} \! \int \! \! \int \! \! dt    dt'  \!  \! & \!  \!  \Theta^k_{P}(t') \! \! & \!
\langle \{{\hat d}(t),{\hat d}(t')\}\rangle \nonumber \\ & \times  \! & \! \! \mathcal{G}^{R,S}_{\hat{{E}}} \left({\bf r}_k(t),t;{\bf r}_k(t'),t'\right). 
\label{eq:phase usual atom interferometry2}
\end{eqnarray}
where the function $\Theta^k_{P}(t')$ is equal to one when the atomic position ${\bf r}_k(t')$ is above the plate and zero elsewhere. Note that this condition automatically bounds the integration domain for the time $t$, since the electric field response function $\mathcal{G}^{R,S}({\bf r}_k(t),t;{\bf r}_k(t'),t')$ yields non-zero values only if the four-position $({\bf r}_k(t),t)$ is on the light cone issued from the image four-position $({\bf r}_{{\rm I}  k }(t'),t')$. Eq.~(\ref{eq:phase usual atom interferometry2}) shows that the single path vdW phase arises from the fluctuating dipole at time $t',$ $d(t'),$
 which  produces an electric
field propagating from its source point ${\bf r}'={\bf r}_k(t')$ to the new atomic position ${\bf r}= {\bf r}_k(t) $  after bouncing off 
the plate (see Fig.~2), where it interacts 
with 
the new atomic dipole $d(t),$  still correlated to the older dipole value $d(t').$
Clearly, the dipole memory time must be larger than the 
time delay corresponding to the round-trip light propagation between atom and surface,
 $\tau= t-t'=  |{\bf r}_k(t'+\tau)-{\bf  r}_{{\rm I}k}(t')| / c \approx 2 z_k(t')/c,$ a condition easily met
in the short-distance limit. Here we model the internal dipole as an harmonic oscillator in order to derive simple analytical
 results:
\(
\langle \{{\hat d}(t),{\hat d}(t')\}\rangle = \omega_0\alpha(0)\cos\left[\omega_0(t-t')\right],
\) 
where $\alpha(0)$ represents the zero-frequency atomic polarizability.

Combining this result with the 
 analytic expression for the field Green's function $\mathcal{G}^{R,S}_{\hat{{E}}}$~\cite{DoublePath}, we derive from 
 (\ref{eq:phase usual atom interferometry2})
($\epsilon_0=$ vacuum permittivity)
\begin{equation}\label{SPinterm}
\varphi_k^{\rm SP}  =  \frac { \omega_0\alpha(0)} {32 \pi \epsilon_0} \int_{0}^T 
 \frac {dt'} {  \overline{z}_k^3(t')}
\end{equation}
The distance $\overline{z}_k(t')= \frac 1 2 \left( z_k(t')+z_k(t'+\tau) \right)$ is
defined in terms of the round-trip propagation time 
$\tau\approx 2z_k(t')/c.$ We neglect second-order terms in the atomic velocity and  assume that the vertical atomic acceleration is
 not exceedingly large ($\ddot{z}_k(t) \ll c^2 / z_k(t) \simeq 5\times 10^{24} {\rm m.s^{-2}}$ for a plate distance $z_k=20\,{\rm nm}$). 
 Using the expression for the vdW potential $V_{\rm vdW}(z)=-\hbar \omega_0 \alpha(0)/(32 \pi\epsilon_0 z^3),$
  the SP phase  (\ref{SPinterm}) 
 is then expressed  directly in terms of 
 the quasi-static standard vdW phase (\ref{eq:instantaneous vdW potential phase}) plus a dynamical first-order correction proportional to the 
 potential gradient $V_{\rm vdW}'(z_k(t)):$
   \begin{eqnarray}
  \label{eq:vdWcoarsegraining}
  \varphi^{\rm SP}_k  =  \varphi^{\rm (vdW)}_k
   - \frac {1} {\hbar}  \int_0^T dt \,
V_{\rm vdW}'(z_k(t))  \frac {z_k(t)  \dot{z}_k(t)} {c}  
  \end{eqnarray}

  \begin{figure}[htbp]
\begin{center}
\includegraphics[width=4cm]{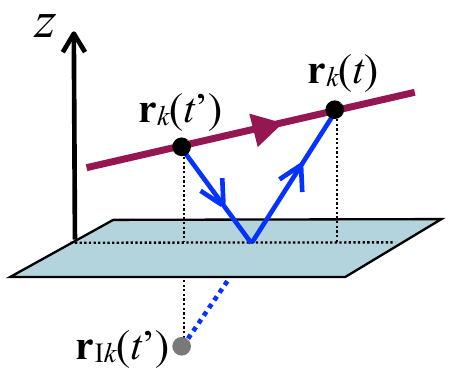}
\end{center}  
\caption{(color online). Effect of retardation on the van der Waals  phase:
the vdW potential as seen by the atomic wave-packet is coarse-grained over the time interval corresponding to the round-trip propagation of light between atom and surface, producing a phase correction proportional to the vertical atomic velocity.} 
\label{fig:round trip}
\end{figure}     
  
  We may also cast the expression (\ref{eq:vdWcoarsegraining}) for $ \varphi^{\rm SP}_k $ in the same form of 
  (\ref{eq:instantaneous vdW potential phase}), provided that we replace the potential taken at the instantaneous position $V_{\rm vdW}(z_k(t)) $ by its 
  coarse-grained  average over the round-trip time $\tau=2 z_k(t)/c:$ $\overline{V}_{{\rm vdW}}(z_k(t))= \frac {1} {\tau} \int_{t}^{t+\tau} dt' V_{\rm vdW}(z_{k}(t'))$. 
  In other words, the position at which the vdW potential is taken cannot be resolved 
  below the scale $\Delta z=2 \dot{z}_k(t)  {z_k(t)}/{c}$ representing the atomic vertical displacement during 
  the round-trip time $2 z_k(t)/c,$ as illustrated by Fig.~2. As far as local phases are concerned, the averaging of the vdW potential 
thus captures the interplay between field retardation and external atomic motion.

 The resulting dynamical correction to the SP phase, given by the second term in the r.-h.-s. of (\ref{eq:vdWcoarsegraining}), turns out to depend on the endpoints only. 
 For the typical non-relativistic velocities employed in atom interferometers, it is a very small phase, smaller than the standard quasi-static vdW phase 
 (\ref{eq:instantaneous vdW potential phase})
 by a factor of the order of $\dot{z}_k/c$ \cite{footnote}.  
It thus seems to be out of experimental reach, for it would be extremely difficult to isolate this phase contribution from the much larger 
 standard vdW phase. 

The
non-local double-path phase contributions $\phi^{\rm DP }_{jk}$ in
(\ref{IF}) are also 
of first-order in $\dot{z}_k/c$ and thus much smaller than the standard vdW phase. However, 
in contrast to the SP phases discussed above, they are non-additive as shown in the following, which
 could be used to isolate them from the  main contribution. 

The physical origin of $\phi^{\rm DP }_{jk}$ is similar to the local phase dynamical correction discussed in connection with
(\ref{eq:phase usual atom interferometry2}), except that it
involves propagation between two different wave-packets.  More precisely, 
$\phi^{\rm DP }_{jk}$ is derived as the difference between the propagation integrals analogous to (\ref{eq:phase usual atom interferometry2}) 
connecting wave-packet $k$ to $j$  and $j$ to $k$ \cite{DoublePath}:
\begin{eqnarray}
& & \phi^{\rm DP }_{jk} =  \frac {1} {4}\int \! \! \int dt   dt'   \Theta^k_{P}(t')
\langle \{{\hat d}(t),{\hat d}(t')\}\rangle \\
  & \times & \,\Bigl[ \mathcal{G}^{R,S}_{\hat{{E}}} \left({\bf r}_j(t),t;{\bf r}_k(t'),t'\right)
 - \mathcal{G}^{R,S}_{\hat{{E}}} \left({\bf r}_k(t),t;{\bf r}_j(t'),t'\right)\Bigr]. \label{eq:DP_general} \nonumber
\end{eqnarray}
where $\Theta^k_{P}(t')$ is the step function previously introduced in Eq.(\ref{eq:phase usual atom interferometry2}). $\phi^{\rm DP }_{jk}$ vanishes in the quasi-static limit, since the two propagation integrals in (\ref{eq:DP_general})
 cancel each other exactly to zeroth-order of $\dot{z}_k/c.$
In contrast with the SP phase (\ref{eq:vdWcoarsegraining}), 
$\phi^{\rm DP }_{jk}$ is thus a pure dynamical phase shift, arising from the 
asymmetry between the propagations from wave-packet $j$ to $k$ and vice-versa, which is brought into play by the finite speed of light and the vertical 
motions of each packet. 

In order to derive an explicit analytical result from (\ref{eq:DP_general}), we assume  that the different atomic paths are in the same vertical plane and share 
the same velocity component parallel to the plate. 
On the other hand, we take arbitrary non-relativistic motions along the perpendicular direction, which correspond to the functions
 $z_k(t),$ under the short-distance condition $\omega_0 z_k/c\ll 1.$ 
 Neglecting as before terms of order $(\dot{z}_k/c)^2,$ 
we derive from (\ref{eq:DP_general})
   \begin{equation}\label{eq:double path phase}
  \phi^{\rm DP}_{jk} = 3\frac{\omega_0\alpha(0)}{4\pi\epsilon_0c} \, \int_0^T dt\, \frac{\dot{z}_k(t)-\dot{z}_j(t)}{(z_j(t)+z_k(t))^3}
  \end{equation}
   Note that this phase is independent of the velocity component parallel to the conductor plane. This follows from translational invariance parallel to the plate
    and from the 
   condition of perfect conductivity. 
   Because it depends linearly on the speed of each trajectory, 
   $  \phi^{\rm DP}_{jk}$ 
    is invariant under time dilatation $z_j\rightarrow {\tilde z}_j(t)\equiv z_j(\Lambda t),$ $j=1,2,$ $T\rightarrow T/\Lambda,$ with $\Lambda$ arbitrary.

  We now stress the main point of this letter: the double-path phase $\phi^{\rm DP}_{jk}$ as given by (\ref{eq:double path phase}) is non-additive, since
 the denominator in  its r.-h.-s. does not allow one to isolate separate contributions from paths $j$ and $k,$ a signature of the non-local nature of $\phi^{\rm DP}_{jk}.$
    This non-additivity is enhanced 
  when considering a geometry for which the third path is much further away from the plate than the first and second paths (see Fig.~1): we take
  $z_3(t) \gg z_1(t),z_2(t)$ and assume that the differences in vertical atomic velocities are of the same order of magnitude $\dot{z}_1(t)-\dot{z}_2(t) \sim \dot{z}_2(t)-\dot{z}_3(t)$.  It then follows from (\ref{eq:double path phase}) that 
  $\phi^{\rm DP}_{13}+  \phi^{\rm DP}_{32} \ll \phi^{\rm DP}_{12}$: the non-additivity  is maximal in this case.
   
   One can actually use the non-additivity in order to isolate the non-local dynamical corrections from the other  phase contributions. 
   For the  three-arm interferometer shown in Fig.~1, we 
   propose to measure separately the three independent
 phase coherences  appearing in 
    Eq.~(\ref{eq:evolved density matrix nonunitary}), 
    $\phi_{jk}\equiv \phi^{(0)}_{jk}+ \frac {1} {\hbar} \mbox{Re} \left[ S_{\rm IF}[\mathbf{r}_j,\mathbf{r}_k] \right] $ with $j,k=1,2,3,$ $j\neq k,$
    by performing interferometric measurements between  the different pairs of arms.   Using (\ref{IF}), we find that 
    the (maximal) violation of  phase additivity gives the desired non-local double-path shift $\phi^{\rm DP}_{12}:$
   \begin{equation}
\phi^{\rm DP}_{12} \approx  \phi_{12}-(\phi_{13}+\phi_{32}) .
   \end{equation} 
  This approach removes all the additive phases, leaving only the non-local dynamical correction to the vdW phase. 
  Thus, the violation of additivity  enables one to isolate the non-local dynamical Casimir phase. This dynamical phase, on the order of a fraction of a micro-radian for realistic experimental parameters~\cite{DoublePath}, cannot be separated from other contribution in an ordinary Mach-Zehnder configuration. 
  In contrast, 
  multiple-beam atom interferometers can be useful to investigate specific dynamical contributions in the Casimir interaction, since they allow one to probe the 
  non-additive nature of the non-local phase contribution.

To conclude, we have shown that the dynamical coupling of a quantum system to an environment can induce non-additive quantum phases. This is a consequence of the non-locality of the influence functional phase, which depends simultaneously on a given pair of distinct quantum paths. This connection has been discussed for external atomic waves coupled to dipole and electromagnetic field quantum fluctuations bounded by a perfect conductor. In this example, the interplay between field retardation effects and the external atomic motion is at the origin of the non-additive influence phases. The dynamical coupling induces both local and non-local relativistic corrections to the standard van der Waals phases.
Local relativistic corrections, associated to individual paths considered separately, take the form of a coarse-graining of the vdW potential. Non-local relativistic corrections are, in contrast, associated to pairs of interferometer paths and cannot be reduced to individual path contributions. Although
of  the same order of magnitude as the local corrections, the non-local relativistic phase corrections are generally non-additive. 
We have proposed a method to isolate them from other phase shifts in a three-path atom interferometer. These results show that the coupling with an environment
 may induce, in addition to decoherence, phase shifts with unusual properties in atom optics.

\acknowledgements

The authors are grateful to Ryan O. Behunin and Reinaldo de Melo e Souza for stimulating discussions. This work was partially funded  by CNRS (France), CNPq, FAPERJ and CAPES (Brasil).


\begin{thebibliography}{25}
\bibitem{footnote}  In addition, our dynamical phase correction is also  smaller than the 
 correction due to the internal atomic dynamics, which corresponds to 
 the time scale $~1/\omega_0.$ The ratio between the external center-of-mass (discussed here) and internal corrections
 is of the order of the inverse ratio between the corresponding time scales: $(\dot{z}_k/c)(\lambda_0/z_k)\ll 1$ ($\lambda_0=2\pi c/\omega_0$ represents
 the typical atomic transition wavelength), which is very small in usual experimental conditions.
\end{thebibliography}
\end{document}